# COVID-19: Forecasting mortality given mobility trend data and non-pharmaceutical interventions

Victor Hugo Grisales Díaz* (1), Oscar Andrés Prado-Rubio (2) and Mark J. Willis (3)

1 - Universidad Libre Seccional Pereira, Facultad Ciencias de la Salud. Grupo de investigación en Microbiología y Biotecnología MICROBIOTEC, Belmonte Avenida Las Américas, Colombia (victorh.grisalesd@unilibre.edu.co)

2 – GIANT, Departamento de Ingeniería Química, Universidad Nacional de Colombia - Sede Manizales, Campus La Nubia, Km 9 vía al Aeropuerto La Nubia, Bloque L, Colombia.

3 - School of Engineering, Newcastle University, Newcastle upon Tyne, NE1 7RU, UK

## Abstract

We develop a novel hybrid epidemiological model and a specific methodology for its calibration to distinguish and assess the impact of mobility restrictions (given by Apple's mobility trends data) from other complementary non-pharmaceutical interventions (NPIs) used to control the spread of COVID-19. Using the calibrated model, we estimate that mobility restrictions contribute to 47 ± 10% (US States) and 47 ± 12% (worldwide) of the overall suppression of the disease transmission rate using data up to 13/08/2020. The forecast capacity of our model was evaluated doing four-weeks ahead predictions. Using data up to 30/06/20 for calibration, the mean absolute percentage error (MAPE) of the prediction of cumulative deceased individuals was 5.0 ± 1.9% for the United States (51 states) and 6.7 ± 2.6% worldwide (49 countries). This MAPE was reduced to 3.5 ± 1.3% for the US and 3.8 ± 1.3% worldwide using data up to 13/08/2020. We find that the MAPE was higher for the total confirmed cases at 11.5 ± 4.7% worldwide and 10.2 ± 2.7% for the US States using data up to 13/08/2020. Our calibrated model achieves an average R-Squared value for cumulative confirmed and deceased cases of 0.992 using data up to 30/06/20 and 0.98 using data up to 13/08/20.

## Highlights

- A compartmental model that incorporates mobility trends data reported by Apple
- The impact of mobility restrictions on the rate of disease suppression is ~47%
- A four week ahead forecast of cumulative deceased cases has a MAPE of ~3.6 %
- A four weeks ahead forecast of cumulative confirmed cases has a MAPE of ~10.6 %

## Introduction

Non-pharmaceutical interventions (NPIs) issued by governments are successfully controlling COVID-19 virus transmission rates. For example, it has been suggested that NPIs have prevented or delayed 61 million new confirmed cases in China, South Korea, Italy, Iran, France, and the United States [1]. Unfortunately, these policies are easing in several countries around the world and consequently global infected cases of COVID-19 are still drastically increasing with the number of daily new infected cases at more than two hundred thousand, (20/09/2020), see [2]. As a result, new NPIs are constantly being issued by governments around the world to reduce rates of contagion.

The introduction of any new NPI needs to be cost-effective, as NPIs have also shown high social and economic costs, such as unemployment. To try to ensure the most cost-effective NPIs are introduced, mathematical models are fundamental as they can be used to understand and measure the impact of NPIs [3]. Furthermore, mathematical models have been used to estimate the effective reproduction number, which is reduced through the application of NPIs [4]. This is an essential metric used by policymakers and epidemiologists as exponential growth in active cases is still observed if this number is higher than one in any determined period. For this reason, in [5,6] we developed a mathematical model that estimated the effective reproduction number as a function of NPIs and time.

This contribution aims to extend our previous work and estimate the dynamics of mortality as well as calculate the cumulative active cases for many countries around the world. At the same time, we evaluate the effectiveness of restrictions on mobility (i.e., walking, driving and transport) on the reduction of the disease transmission rate and hence the control of the cumulative number of infected and deceased individuals. To do this we use mobility trends data provided by Apple [7]. This is important as mobility restrictions have been identified as being the most significant of all NPIs by [8,9]. For example, it has been estimated that lockdowns in the US would save 1.7 million lives by October 2020 with a monetary mortality benefit of 8 trillion USD [10]. In addition, country-wide lockdowns have been found to reduce disease transmission rates by 75–87% in 11 European countries [8]. Similarly, school closure, quarantine, and distance working have been estimated to have reduced the number of contagious individuals by 78.2%-99.3% [11]. However, there are also reports of the lower effectiveness (a 47% reduction in contagion rates) being associated with mobility [9], which demonstrates the wide confidence bounds associated with the evaluation of the impact of this NPI.

In this work, a new compartmental model is proposed and calibrated to distinguish and measure the impact of mobility restrictions on the rate of disease suppression. In addition, we use the calibrated model to forecast up to four-weeks ahead the number of cumulative infected cases and the resulting mortality rates. Here, we demonstrate for the first time for a large number of cases (to the best of our knowledge) that including both mobility trend data and NPIs are necessary to capture dynamics and simultaneously accurately forecast mortality. As we use a hybrid compartmental model, the mathematical model developed here has the additional advantage that can be also used to calculate the effective reproductive number at any particular time and for any country. To generate our results we use data from 49 countries and 51 U.S states available from [12] and [2], respectively.

## Methods

In this contribution, our previous model [5] is extended to predict mortality and to include a term to estimate the reduction on the contagious rates given reported mobility data. Due to these modifications, the proposed model is referred to as SIRD-MC (mobility-control).

## SIRD-MC model

To develop the SIRD-MC model, individuals are compartmentalised into the number of susceptible ($S(t)$) compared to the total population ($N$), the number of infected ($I(t)$), the number of removed ($R(t)$), and the number of deceased ($D(t)$). The compartmental $R(t)$ is the summary of recovered and deceased individuals. We also include a term for patients that have symptoms that can lead to death ($I_d(t)$). This compartment is proposed to deal with different kinds of infected individuals e.g. asymptomatic and symptomatic, as it has been shown that this is necessary in order to analyse the rates of contagion [13]. Our SIRD-MC model consists of the following ordinary differential equations,

$$\frac{dI(t)}{dt} = \beta(t) \cdot I(t) \cdot \frac{S(t)}{N} - \alpha \cdot I(t) \tag{1}$$

$$\frac{dS(t)}{dt} = -\beta(t) \cdot I(t) \cdot \frac{S(t)}{N} \tag{2}$$

$$\frac{dR(t)}{dt} = \alpha \cdot I(t) \cdot x \tag{3}$$

$$\frac{dI_d(t)}{dt} = y \cdot \beta(t) \cdot I_d(t) \cdot \frac{S(t)}{N} - \alpha \cdot I_d(t) \tag{4}$$

$$\frac{dD(t)}{dt} = \mu \cdot I_d(t) \tag{5}$$

$$C(t) = I(t) + R(t) \tag{6}$$

In the above equations, the removal rate of reported infectious individuals (or recovery rate), $\gamma$ ($\text{day}^{-1}$), is assumed constant and equal to $1/8 \, \text{day}^{-1}$, as it has been shown to be reasonable and effective [5,9]. The transmission rate of the virus ($\beta(t)$) is assumed to be dependent on NPIs (this is described in detail in the next section). The model also includes an under-reported parameter ($x$) for removed individuals [5] and we assume that the rate constant associated with $I_d(t)$ is proportional to that of ($I(t)$) where ($y$) is an estimated constant between zero and one. As only the compartment $I_d(t)$ within our model can lead to death, the deceased individuals, $D(t)$, are considered to be dependent of $I_d(t)$. The total deceased individuals are modelled assuming a constant fatality rate $\mu$, $(\text{day}^{-1})$. Finally, the total confirmed cumulative cases ($C(t)$) are modelled by summing the infected and removed cases ($R(t)$).

### Non-Pharmaceutical Intervention policies

In our previous work, we introduced a time-varying term that accounts for a reduction in the virus transmission rate constant, $\beta(t)$, using a first-order differential equation employing all NPIs as input [5]. We assumed a control signal $\bar{u}(t)$ which was a Boolean variable, i.e. it could take the value of zero (NPIs off) or one (NPIs are active), and the time varying disease transmission rate $\beta(t)$ was assessed through determination of the model parameters associated with the ODE model. In this work, we modify this expression to,

$$\tau \cdot \frac{d\beta'(t)}{dt} + \beta'(\text{t}) = -k_a \cdot \bar{u}(t)_a - k_m \cdot \bar{u}(t)_m \qquad (7)$$
$$\beta(t) = \beta_I + \beta'(t)$$

In this equation, we have the rate of reduction of the disease transmission rate ($\beta'(t)$) which is assumed zero at the initial time with an initial rate ($\beta_I$). In addition, $\tau$ represents a time constant associated with the reduction in the transmission rate, i.e. it is a measure of the time required for the full reduction in the disease transmission rate to occur as a result of the application of any NPIs. The constants $k_m$ and $k_a$ describe the reduction in the rate of disease transmission because of a reduction in mobility and all other NPIs. The signal $\bar{u}(t)_m$ represents the effective reduction in mobility (as suggested by the Apple *Mobility Trends Reports*) because of a lockdown or other measures. While the signal $\bar{u}(t)_a$ represents the application of all other NPIs, such as the use of masks, biosecurity protocols, or closure of schools. By including these two terms in the model, we can capture the effects of disease suppression with and without restrictions on mobility.

Given the estimated numerical value of these two model parameters, we can estimate the contribution of mobility restrictions to the overall reduction in disease transmission rate as,

$$S_m = \frac{\bar{k}_m}{\bar{k}_a + \bar{k}_m} \qquad (8)$$

In this equation, $\bar{k}_m$ is the average $k_m$ for all US states or countries evaluated and similarly, $\bar{k}_a$ is the average value of $k_a$.

### Mobility trend - modelling and prediction

*Mobility Trends Reports* provided by Apple were used as inputs of the model as the data is easily accessible, updated and conveniently normalised [7]. Mobility trends for 49 countries and 51 U.S. states were averaged for each day using all reported mobility data (i.e. walking, driving, transport), converting the data into a representative number per day. Firstly, the data is normalised to have values between zero (no mobility restrictions) and one (maximum mobility restriction). Subsequently, we transform this data into a mobility signal ($u_m$) using linear regression. To model the mobility signal, we consider this signal to be linear and have five changes in slope. See Appendix A for more details of this mathematical modelling.

### Refined approach for model parameters calibration

To demonstrate the prediction accuracy of the SIRD-MC model, the model parameters were tuned separately for 49 countries and all US states (we calibrate a model for a particular country provided Apple mobility data is available and that the particular country publishes the number of recovered cases of individuals with COVID-19). The seven model parameters, $k_m$, $k_a$, $x$, $\beta_o$, $\mu$, $\tau$ and $y$ are determined such that a least squares type objective function is minimised (see Appendix B) and as our implementation is in MATLAB we use the built in function 'lsqnonlin' available in MATLAB's optimisation toolbox. Subsequently, parameter confidence intervals were calculated using the additional function 'nlparci'. The estimation of the parameter confidence intervals of this method is based on the covariance matrix and t-student test for normal distribution.

### Model Forecasting predictive power – model validation

Once the parameters of the kinetic model are fitted, the predictive capability of the model needs to be validated [14,15]. To validate our model, we made predictions up to four weeks ahead of the cumulative deceased individuals and number of active cases. For forecasting of the total cases, the initial conditions of the ODEs were specified as $I_d(t_2)$, $I(t_2)$ and $\beta(t_2)$, and the initial condition for recovered and deceased individuals set as,

$$R(t_2)= I_R(t_2)+R_R(t_2)\text{-}I(t_2) \tag{9}$$

$$D(t_2)= D_R(t_2) \tag{10}$$

where $t_2$ is equal to 30 Jun 2020 as we used data for model calibration up to this date. To compare predictive capacity of our model in relation to the number of data, we use data up to 13 Aug 2020 ($t_2$) for model calibration. To estimate the confidence bounds in the predictions, we used the same initial conditions described above.

The median absolute percentage error (MdAPE) and the mean absolute percentage error (MAPE) were calculated to report the predictive capability of our models. Usually, the MAPE is estimated using an absolute percentage error (APE) calculated with the mean value of the prediction as reference, however, when calculating it in this way, it is known that the MAPE is not robust against outliers [16]. Here, as an outlier-resistant measure, we estimated the APE three times for each experimental point using the three simulated values of a prediction (the mean value of the prediction and its respective maximum and minimum bounds). The lowest APE of the three is then selected for each experimental point to estimate the MAPE. By doing this, we can calculate the most likely MAPE avoiding potential errors due to outliers.

## Results

### Mobility trend model and forecasting

To perform model calibration and the subsequent validation, the first step is to tune the mobility model, which defines the SIRD-MC model input signal. This model is also required to perform the forecasting of mobility for model validation. The mobility restriction model consists of a linear representation where the slope of the curve can potentially change over five windows of time (see Appendix A). As an example, in **Figure 1** we show the calibrated model and data up to 29 Jun 2020 (blue) and forecasting of future mobility (data in black and model in light blue) up to four weeks ahead for Brazil, Italy, Australia, and Colombia.

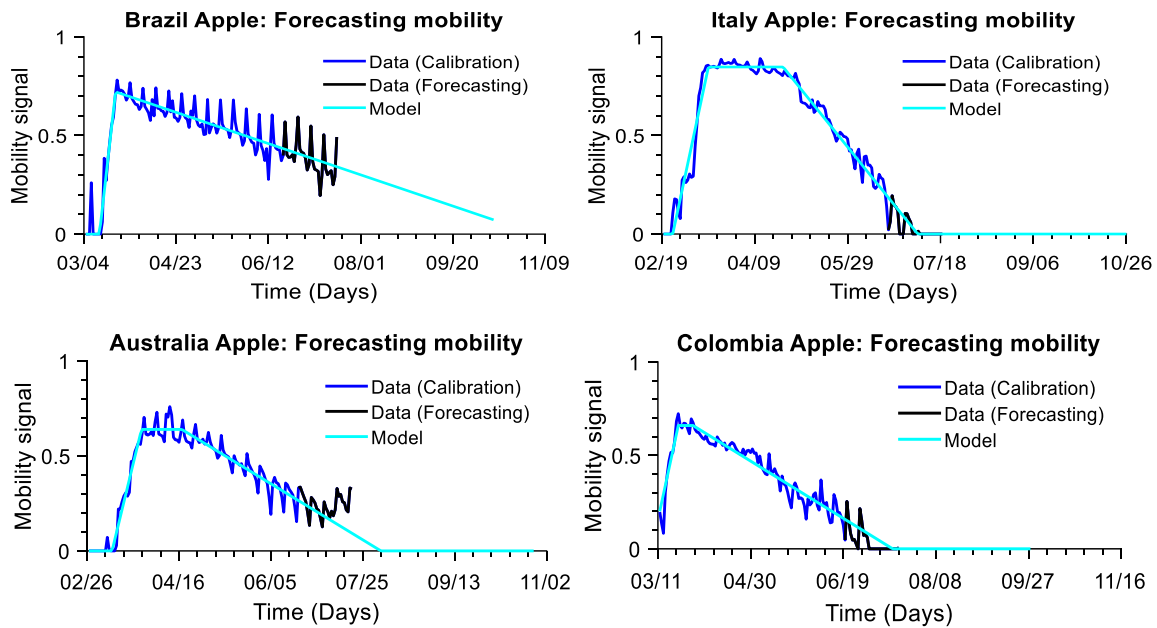

*Figure 1. Mobility data from apple and control signal estimated (inverse of the Mobility*

A reduction of the predictive capacity of the model is observed when new and strong mobility restrictions are introduced. As an example, new lockdown rules in Melbourne (Australia) were issued to stop the spread of the virus, and the trend is beyond the predictive power of our model (see **Figure 1**). For this reason, we limited our study to make short term predictions of up to four weeks. The average predictive power of the mobility model in all cases studied is shown in Figure 2. Although the raw data is still highly noisy, the mean absolute error of the model was lower than 0.06. Moreover, it may be observed that the forecasting error is lower than the calibration error as 37 of the 49 countries studied does not have restrictions of mobility. This is important for parameter identification as the mobility signal covers all the spectrum of mobility, i.e. the effect of mobility on the reduction of the contagious rates is observed in the data. Although mobility tendencies provided by Apple (i.e. walking, driving, transport) are back to normal for most of the countries around the globe, some of the countries still have a reduction in mobility (22/07/2020) such as South Korea, Brazil, Chile, Argentina, South Africa, Australia, amongst others.

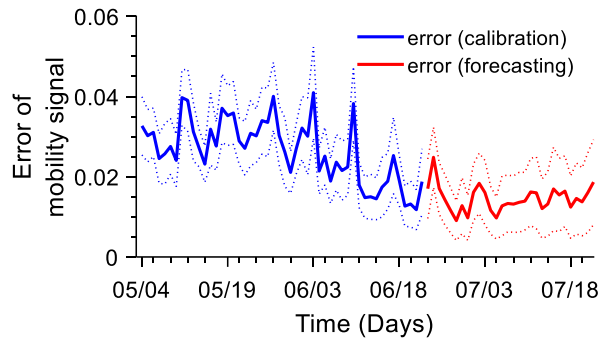

*Figure 2. A worldwide error of the mobility restriction model for data reproduction and forecast. Dotted lines are the 95% confidence bounds*

### Disease suppression with and without mobility restrictions

Considering the countries studied in this work and data up to 30 Jun 2020, the estimated gains (i.e. $k_m, k_a$) of the ODE (7) were evaluated. In the 49 countries studied, the gain associated with mobility restrictions ($k_m$) was on average 0.183 ± 0.054 ($\bar{k}_m$) and the gain associated with all other NPIs ($\bar{k}_a$) was equal to 0.2050 ± 0.03 (see Table 1), i.e. worldwide mobility restrictions contribute to 47±14% of the overall reduction in the disease transmission rate of COVID-19. Whereas for the US States, mobility restrictions account for 54±14% of the overall reduction in the disease transmission rate. When data up to 13/08/2020 was used for calibration, our models indicate that mobility restrictions account for 47 ± 10% (US States) and 47 ± 12% (worldwide) of the overall reduction in the disease transmission rate. The estimated model parameters for each country and the US States are reported in Appendix C.

**Table 1.** Average parameters of SIRD-MC for the States of the United States and 49 countries

| Case | Parameters | | | | | | |
|---|---|---|---|---|---|---|---|
| | $\bar{k}_a$ | $\bar{x}$ | $\bar{\beta}_o$ | $\bar{k}_m$ | $\bar{\mu}$ | $\bar{\tau}$ | $\bar{y}$ |
| Data up to 30/06/2020 | | | | | | | |
| Worldwide | 0.205±0.03 | 41.13±7.23 | 0.3742±0.025 | 0.183 ± 0.054 | 0.3755±0.296 | 28.58±4.16 | 0.8944±0.0253 |
| U.S. States | 0.254±0.02 | 28.37±6.28 | 0.402±0.021 | 0.303±0.075 | 0.357±0.21 | 31.0421±3.9 | 0.8317±0.0201 |
| Data up to 13/08/2020 | | | | | | | |
| Worldwide | 0.208±0.03 | 39.4±7.23 | 0.3597±0.028 | 0.1863±0.048 | 0.2284±0.185 | 35.06±5.99 | 0.9026±0.0222 |
| U.S. States | 0.264±0.03 | 25.6±6.47 | 0.417±0.031 | 0.231±0.0495 | 0.1785±0.0954 | 34.02±6.5 | 0.8803±0.0208 |

In Table 1 we show all the model parameters – averaged over all countries and the US states. Note that the time constant ($\bar{\tau}$) represents the time necessary to achieve 63.2% of the total change in contagious rates. Worldwide, the time constant is 28.58 ± 4.16 (days), data up to 30/06/2020. The parameter with the highest variation was the death rate for COVID-19 patients with symptoms ($\bar{\mu}$) as a small reduction in the parameter '$\gamma$' can lead to a high reduction in $\mu$, i.e. '$\gamma$' has an exponential effect on the number of patients with symptoms.

The initial or basic reproductive number of the epidemic was found to be on average 3.0 ± 0.2 (worldwide) and 3.2 ± 0.17 (US States). Note, the low confidence bounds of our average basic reproduction number. In contrast, the average basic reproduction number on 11 European countries was estimated by [8] as 3.8. We estimated that the initial reproduction number was drastically reduced due to NPIs (the initial number is ~3). However, most of the countries and the US states up to 13 Aug 2020 has shown an effective reproductive number higher than one, see Fig. 3. As confinement policies were found to be significant to control the effective reproduction number in most of the countries, it is evident the need of social distancing in countries with high transmission rates to minimise the number of deaths.

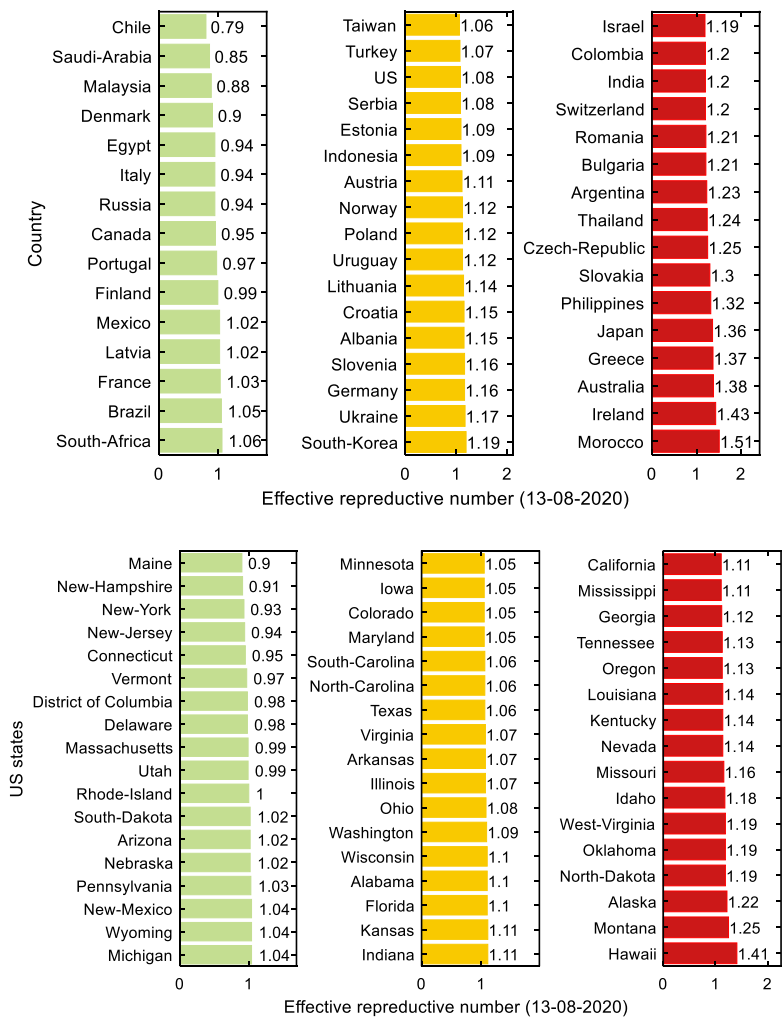

Figure 3. Effective reproductive number of COVID-19 on US states and worldwide

### Average forecasting capacity of cumulative deaths and confirmed cases

We have validated our model doing forecasting up to four weeks ahead of the total confirmed cases and deaths. For illustration purposes, the forecasting of the cumulative cases and cumulative number of diseased individuals for several countries are shown in Figure 4.

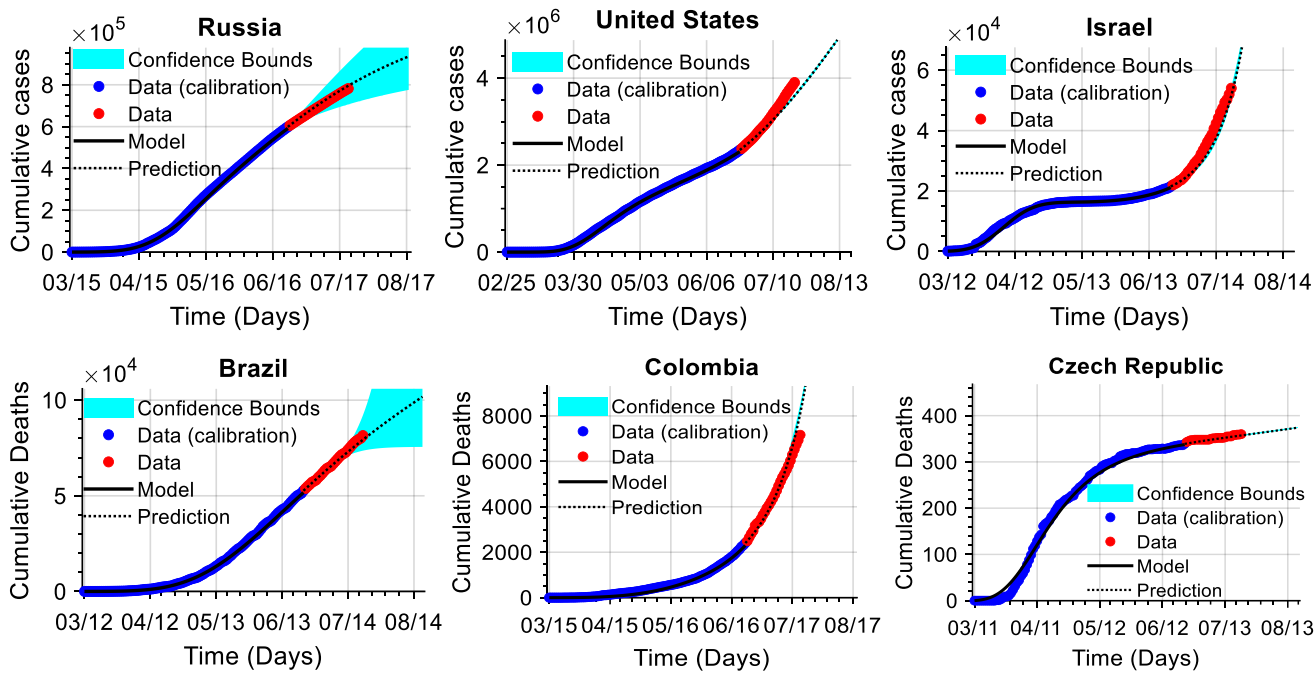

*Figure 4. Profiles of cumulative deceased individuals confirmed cases of COVID-19 up to 28/07/2020. Red data were not used on model calibration*

As expected, the average MAPE for US States and worldwide increases with the number of days ahead (see Figure 5 and Figure 6). In all cases considered, the MAPE for two weeks ahead was 2.4 ± 0.8% for cumulative deceased individuals (Figure 5) and 3.9 ± 0.9% (Figure 6) for cumulative confirmed cases. When the number of weeks ahead was increased from two to four, the mean MAPE was increased by 2.5- and 3.5-fold for deceased individuals and confirmed cases. The MAPE for cumulative deaths was lower than 10% in 41 U.S. states and 39 countries. Hence, for the forecasting of cumulative deaths, our model has an 80% probability of producing four-week ahead predictions with a mean MAPE lower than 10%. Whereas, if we consider the confirmed cases, this probability is 62%. Using all data (calibration and validation), our model had a R-Squared value of 0.992 (calibration data up to 30/06/2020) and 0.98 (calibration data up to 13/08/28).

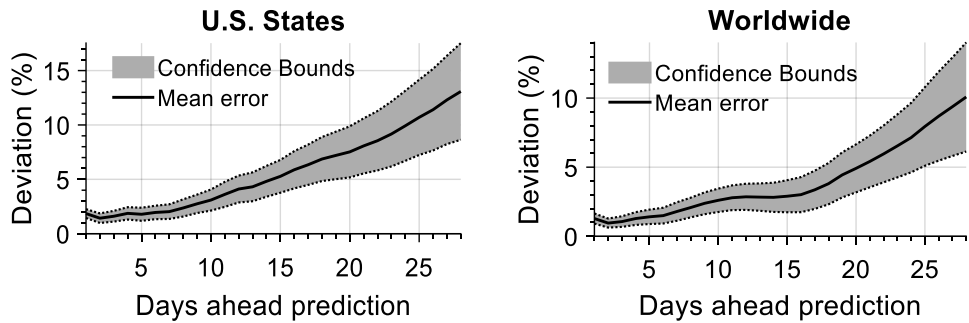

*Figure 5. Forecasting capacity of confirmed cases from 30/06/2020*

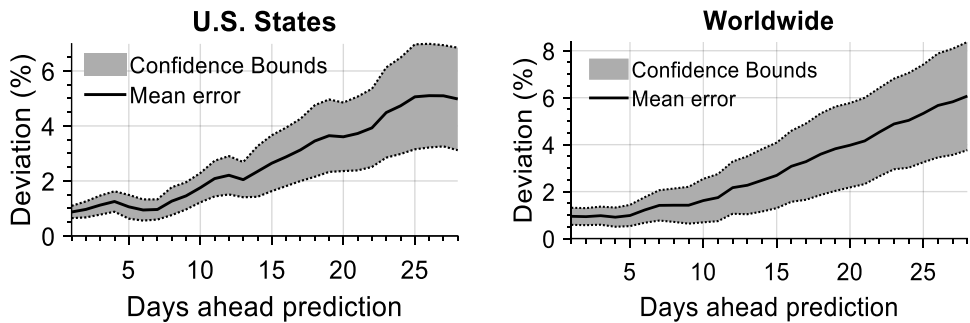

*Figure 6. Forecasting of cumulative deceased individuals from 30/06/2020*

### A comparison of the forecasting capability of the model with previous work

Among the epidemiological models used for forecasting in the literature, the models with lower MdAPE for four weeks ahead prediction of the cumulative number of deaths are the hybrid mortality spline and epidemiological compartment model (IHME-MS SEIR model) and Youyang Gu model (MdAPE ~6.5%) [17], see Table 2. These models were evaluated using 156 and 73 countries, respectively.

**Table 2.** Comparison of models of forecasting for 28 days-ahead for cumulative deaths

| Model | Number of countries and U.S. states | MdAPE | |
|---|---|---|---|
| | | Deaths | Cases |
| IHME-MS SEIR model [17] | 156* | 6.4% | Not estimated |
| Youyang Gu - SEIR [18] | 73* | 6.5% | Not estimated |
| LANL–Growthrate - Dynamic Growth [19] | 131* | 8.0% | Not estimated |
| SIRD-MC (This work data up to 30/06/2020) | 100 | 5.9% | 13.6% |
| | 51 | 5.3% (U.S. States) | 15.8% (U.S. States) |
| | 49 | 6.5% (countries) | 6.4% (countries) |
| SIRD-MC (This work data up to 13/08/2020) | 100 | 2.8% | 11.2% |
| | 51 | 2.8% (U.S. States) | 17% (U.S. States) |
| | 49 | 2.5% (countries) | 8.4% (countries) |

*The MdAPE was estimated and reported by IHME [17]

As in [17], MdAPE is estimated instead of MAPE, we also calculate this figure for comparative purposes. In this work, the MdAPE of SIRD-MC for all states was 5.9% (51 U.S. states and 49 countries worldwide). To estimate the predictive capacity once more data is gathered, we also perform forecasting using a more recently updated dataset (calibration using data up to 13/08/2020, see Table 2). As expected given the increased amount of data, our model increases the predictive capacity with time as both MdAPE for deceased individuals confirmed cumulative cases reduces from 5.9 to 2.8 % and 13.6 to 11.2 %, respectively.

## Discussion and Conclusions

Mobility restrictions imposed by governments are controversial as individual freedom is compromised for the greater good. Lockdowns slow down the spread of the virus; this is a fact. However, the impact of mobility restrictions on the control of the spread of COVID-19 that has been reported in the literature vary in a wide range, between 47 and 99% [9]. Here, using data up to 30 Jun 2020 and our model (SIRD-MC), we demonstrated that mobility reductions could explain ~47 and ~54% of the reduction in the disease transmission rates in 49 countries and the US States, respectively. These results are similar to those obtained by [7], where data from 26 countries was used to calibrate a hybrid model combining the susceptible- infected-recovered (SIR) differential equations with gradient boosted trees (GBT).

An augmented version of the SEIR compartmental model that uses Apple's mobility trend data was proposed by [20] to estimate mortality and hospitalizations. In their work, they demonstrate that mobility data provides a leading indicator of contagious rates in eight US States. However, their model assumed that the change in the value of the effective reproduction number only depended on the reduction of mobility. In our work, we demonstrate that this is not true as other NPIs were found to reduce the spread of the virus by around ~46 and 53%, approximately.

To validate the predictive capacity of our model we considered the forecasting of cumulative active cases as well as the numbers of diseased individuals up to four weeks ahead and provide several statistical indicators of the forecasting accuracy such as MAPE, MdAPE and R-Squared. We found that our model has a lower MdAPE than other models studied by [17]. At the same time, our calibrated model was able to capture the disease suppression dynamics with a high R-Squared, 0.992 (calibration data up to 30/06/2020). Hence, we believe that the modelling approach suggested in this paper to estimate mortality will be a useful tool that can assist decision making processes.

## Competing interests

The authors declare that they have no competing interests.

## Acknowledgement

This research did not receive any specific grant from funding agencies in the public, commercial, or not-for-profit sectors.

## References

[1]  S. Hsiang, D. Allen, S. Annan-Phan, K. Bell, I. Bolliger, T. Chong, H. Druckenmiller, L.Y. Huang, A. Hultgren, E. Krasovich, P. Lau, J. Lee, J. Rolf, J. Tseng, T. Wu, The effect of large-scale anti-contagion policies on the COVID-19 pandemic., Nature. (2020). https://doi.org/10.1038/s41586-020-2404-8.

[2]  Wolrdometers, Coronavirus Cases, (2020). https://doi.org/https://www.worldometers.info/coronavirus/country/us/.

[3]  C.O. Buckee, M.A. Johansson, Individual model forecasts can be misleading, but together they are useful, Eur. J. Epidemiol. (2020) 731–732. https://doi.org/10.1007/s10654-020-00667-8.

[4]  Y. Liu, A.A. Gayle, A. Wilder-Smith, J. Rocklöv, The reproductive number of COVID-19 is higher compared to SARS coronavirus, J. Travel Med. 27 (2020) 1–4. https://doi.org/10.1093/jtm/taaa021.

[5]  M.J. Willis, V.H.G. Díaz, O.A. Prado-Rubio, M. von Stosch, Insights into the dynamics and control of COVID-19 infection rates, Chaos, Solitons & Fractals. 138 (2020) 109937. https://doi.org/10.1016/j.chaos.2020.109937.

[6]  M.J. Willis, A.R. Wright, V. Bramfitt, V.H. Grisales Díaz, COVID-19: Mechanistic model calibration subject to active and varying non-pharmaceutical interventions, MedRxiv. (n.d.). https://doi.org/https://doi.org/10.1101/2020.09.10.20191817.

[7]  Apple, Apple Mobility Trends Reports, (2020). https://covid19.apple.com/mobility.

[8]  S. Flaxman, S. Mishra, A. Gandy, H.J.T. Unwin, T.A. Mellan, H. Coupland, C. Whittaker, H. Zhu, T. Berah, J.W. Eaton, M. Monod, P.N. Perez-Guzman, N. Schmit, L. Cilloni, K.E.C. Ainslie, M. Baguelin, A. Boonyasiri, O. Boyd, L. Cattarino, L. V. Cooper, Z. Cucunubá, G. Cuomo-Dannenburg, A. Dighe, B. Djaafara, I. Dorigatti, S.L. van Elsland, R.G. FitzJohn, K.A.M. Gaythorpe, L. Geidelberg, N.C. Grassly, W.D. Green, T. Hallett, A. Hamlet, W. Hinsley, B. Jeffrey, E. Knock, D.J. Laydon, G. Nedjati-Gilani, P. Nouvellet, K. V. Parag, I. Siveroni, H.A. Thompson, R. Verity, E. Volz, C.E. Walters, H. Wang, Y. Wang, O.J. Watson, P. Winskill, X. Xi, P.G. Walker, A.C. Ghani, C.A. Donnelly, S.M. Riley, M.A.C. Vollmer, N.M. Ferguson, L.C. Okell, S. Bhatt, Estimating the effects of non-pharmaceutical interventions on COVID-19 in Europe, Nature. 584 (2020) 257–261. https://doi.org/10.1038/s41586-020-2405-7.

[9]  D. Delen, E. Eryarsoy, B. Davazdahemami, No Place Like Home: Cross-National Data Analysis of the Efficacy of Social Distancing During the COVID-19 Pandemic, JMIR Public Heal. Surveill. 6 (2020) e19862. https://doi.org/10.2196/19862.

[10]  M. Greenstone, V. Nigam, Does Social Distancing Matter?, SSRN Electron. J. (2020). https://doi.org/10.2139/ssrn.3561244.

[11]  J.R. Koo, A.R. Cook, M. Park, Y. Sun, H. Sun, J.T. Lim, C. Tam, B.L. Dickens, Interventions to mitigate early spread of SARS-CoV-2 in Singapore: a modelling study, Lancet Infect. Dis. 20 (2020) 678–688.

https://doi.org/10.1016/S1473-3099(20)30162-6.

[12] C.-19 Map, Johns Hopkins Coronavirus Resource Center, (2020). https://coronavirus.jhu.edu/map.htm.

[13] M. Chen, M. Li, Y. Hao, Z. Liu, L. Hu, L. Wang, The introduction of population migration to SEIAR for COVID-19 epidemic modeling with an efficient intervention strategy, Inf. Fusion. 64 (2020) 252–258. https://doi.org/10.1016/j.inffus.2020.08.002.

[14] J. Almquist, M. Cvijovic, V. Hatzimanikatis, J. Nielsen, M. Jirstrand, Kinetic models in industrial biotechnology – Improving cell factory performance, Metab. Eng. 24 (2014) 38–60. https://doi.org/10.1016/j.ymben.2014.03.007.

[15] P. Nadella, A. Swaminathan, S. V. Subramanian, Forecasting efforts from prior epidemics and COVID-19 predictions, Eur. J. Epidemiol. (2020) 727–729. https://doi.org/10.1007/s10654-020-00661-0.

[16] L.M. Basson, P.J. Kilbourn, J. Walters, Forecast accuracy in demand planning: A fast-moving consumer goods case study, J. Transp. Supply Chain Manag. 13 (2019) 1–9. https://doi.org/10.4102/jtscm.v13i0.427.

[17] J. Friedman, P. Liu, E. Gakidou, * IHME COVID-19 Model Comparison Team, Predictive performance of international COVID-19 mortality forecasting models, MedRxiv. (2020). https://doi.org/https://doi.org/10.1101/2020.07.13.20151233.

[18] Y. Gu, COVID-19 Projections Using Machine Learning, (2020). https://covid19-projections.com/.

[19] Los Alamos Natinoal Laboratory COVID-19 Confirmed and Forecasted Case Data, (2020). https://covid-19.bsvgateway.org/.

[20] A.C. Miller, N.J. Foti, J.A. Lewnard, N.P. Jewell, C. Guestrin, E.B. Fox, Mobility trends provide a leading indicator of changes in SARS-CoV-2 transmission, MedRxiv. (2020) 2020.05.07.20094441. https://doi.org/10.1101/2020.05.07.20094441.